\documentclass[prd,aps,nofootinbib,notitlepage,preprintnumbers
]{revtex4}
\usepackage{graphicx,epsf,amsmat
h,amsfonts,amssymb,amsbsy}
\usepackage{epsfig}
\newcommand{\mat}{\left ( \begin{array}}
\newcommand{\emat}{\end{array} \right )}
\newcommand{\vect}{\left ( \begin{array}{c}}
\newcommand{\evect}{\end{array} \right )}

\def\cP{\mathcal P}
\def\cT{\mathcal T}

\begin{document}


\title{
 Spontaneous non-Hermiticity in the (2+1)-dimensional Gross-Neveu model}
\author{T. G. Khunjua $^{1}$, K. G. Klimenko $^{2}$, and R. N. Zhokhov $^{2,3}$ }

\affiliation{$^{1}$ The University of Georgia, GE-0171 Tbilisi, Georgia}
\affiliation{$^{2}$ State Research Center
of Russian Federation -- Institute for High Energy Physics,
NRC "Kurchatov Institute", 142281 Protvino, Moscow Region, Russia}
\affiliation{$^{3}$  Pushkov Institute of Terrestrial Magnetism, Ionosphere and Radiowave Propagation (IZMIRAN),
108840 Troitsk, Moscow, Russia}

\begin{abstract}
Using a nonperturbative approach based on the Cornwall-Jackiw-Tomboulis effective action 
$\Gamma(S)$ for composite operators ($S$ is the full fermion propagator), 
the phase structure of the simplest massless (2 + 1)-dimensional Gross-Neveu model is 
investigated. We have calculated $\Gamma(S)$ and its stationary (or Dyson-Schwinger)
equation in the first order of the bare coupling constant $G$
and have shown that there exist a well-defined dependence of $G\equiv G(\Lambda)$ on the cutoff 
parameter $\Lambda$, such that the Dyson-Schwinger equation is renormalized. It has three different 
solutions for fermion propagator $S$ corresponding to possible dynamical appearance of three 
different mass terms in the model. One is a Hermitian, but two others are non-Hermitian and 
$\cP\cT$ even or odd. It means that 
two phases with spontaneous non-Hermiticity can be emerged in the system. Moreover, mass spectrum of 
quasiparticles is real in these non-Hermitian and $\cP\cT$ even/odd phases.
\end{abstract}
\maketitle

\section{Introduction}

For a long time, it was believed that to describe quantum systems it is necessary to use theories 
with Hermitian Hamiltonians (or Lagrangians), since in this case the energy spectrum is real.
However, in recent decades, it has been understood that there are situations, especially in open 
physical systems interacting with the environment, which can be effectively considered in the 
framework of non-Hermitian Hamiltonians (see, e.g., in review \cite{Ashida}). Moreover, it was 
claimed that if non-Hermitian theories have in addition a space-time reflection 
symmetry $\cP\cT$, then its energy spectrum is real \cite{Bender,Bender2}, i.e. the Hermitian 
nature of the Hamiltonian is only a sufficient, but far from necessary, condition for the 
energy spectrum of the system to be real. This statement is supported by a number of bosonic theories
in which non-Hermiticity together with $\cP\cT$ symmetry lead to real mass spectrum. 

However, the situation is more involved in fermionic systems. On the one hand, indeed, as the 
considerations of some (1 + 1)- and (3 + 1)-dimensional (D) and non-Hermitian field theory models 
with four-fermion interaction show, the $\cP\cT$-symmetry together with non-Hermiticity 
leads to a real spectrum of mass \cite{Klevansky,Felski,Felski2}. On the other hand, in the same 
paper \cite{Felski2} other non-Hermitian and anti-$\cP\cT$-symmetric extensions of the four-fermion 
models are also presented, in which, nevertheless, a real spectrum of fermion masses is also 
generated, i.e. in fact $\cP\cT$ symmetry of the model is not a necessary condition for real fermion 
masses to exist. Thus, the relationship between the phenomena of non-Hermiticity, $\cP\cT$ symmetry 
and the reality of the energy spectrum in any quantum system remains a far from solved problem and 
deserves further study. Moreover, it should be noted one more feature of the non-Hermiticity 
phenomenon, which was observed recently just within the framework of the (3+1)-D Nambu--Jona-Lasinio (NJL) 
model \cite{NJL} with four-fermion interactions. Namely, in this model the non-Hermiticity can arise 
spontaneously \cite{Chernodub2}. (Quite recently, it was noted in Ref. \cite{Mavromatos} that, perhaps, the phenomenon of spontaneous non-Hermiticity occurs also in some models of Yukawa type.) It means that (i) initial Lagrangian of the massless NJL model is 
taken to be Hermitian and $\cP\cT$-symmetric, (ii) but, as it was proved in Ref. \cite{Chernodub2},  
there exists a ground state corresponding to a dynamical (spontaneous) appearance of the 
$\cP\cT$-symmetric and non-Hermitian Yukawa-type term in the effective Lagrangian. In addition, quasiparticle 
excitations of this ground state obey a real mass.

In the present paper, we show that in the simplest (2+1)-D massless Gross-Neveu (GN) model (for 
the first time, it was discussed in Ref. \cite{GN}) with four-fermion interaction
(initially, its Lagrangian is Hermitian and $\cP\cT$-symmetric \footnote{Its Lagrangian is presented 
below in Eq. (\ref{n1}). }) the non-Hermiticity 
can also arise spontaneously. In contrast to the situation observed in the massless NJL model, 
in (2+1)-D massless GN model (i) both $\cP\cT$- and anti-$\cP\cT$-symmetric non-Hermitian ground 
states are allowed to be realised. And (ii) it means that mass terms, and not terms with Yukawa
interaction, with corresponding symmetry properties are generated in the model Lagrangian. 

In this connection, it is necessary to note that recently much attention has been paid to the 
investigation of (2 + 1)-D field theory models, which can be used to predict and study the condensed 
matter physical phenomena of planar nature such as quantum Hall effect, high-temperature 
superconductivity, low-energy graphene physics, etc. To a fairly large extent, these phenomena are 
usually considered within the framework of models with a four-fermion interaction 
\cite{Semenoff,Shovkovy,Gusynin,Mesterhazy,Vshivtsev,Khudyakov,Kanazawa,Gomes,Ebert}. 
One of the reasons is that in these models the spontaneous symmetry breaking occurs dynamically, 
i.e. without taking into account additional scalar Higgs bosons. Moreover, despite the perturbative 
nonrenormalizability of these (2 + 1)-dimensional models, in the framework of nonperturbative 
approaches such as large-$N$ technique, etc., they are renormalizable \cite{Rosenstein}. 
And just using this nonperturbative $1/N$ approach, spontaneous symmetry breaking and the 
associated effect of dynamical generation of the fermion mass were investigated in the simplest 
(2 + 1)-D GN model with four-fermion interaction. 
In particular, it was shown, e.g., in Ref. \cite{Modugno} that at zero temperature and zero
chemical potential (as well as at fixed value of the cutoff parameter $\Lambda$) in this 
(2+1)-D GN model a phase with dynamical chiral symmetry breaking occurs only at sufficiently 
large (positive) values of the bare coupling constant $G\equiv G(\Lambda)$. 
For a rather weak interaction, the symmetric phase is realized in the model, and it is not an 
asymptotically free one. (In contrast, the (1+1)-D GN model \cite{GN} is an asymptotically free and 
dynamic generation of the fermionic mass occurs there for arbitrary values of bare coupling 
constant.) Qualitatively the same properties of this (2+1)-D GN model one can observe in terms 
of variational optimized expansion technique \cite{Klimenko:1993} and other nonperturbative 
variational approarches \cite{Kneur:2007}, etc.

Unlike the aforementioned papers, we investigate phase structure of the (2+1)-D GN model (\ref{n1}) 
within the framework of another nonperturbative approach based on the effective action for 
composite operators. Originally, 
the approach was proposed in the paper by Cornwall-Jackiw-Tomboulis (CJT) \cite{CJT} when 
considering mainly a scalar $\phi^4$-field model, etc. Then in a series of papers 
\cite{Peskin,Casalbuoni,Dorey,Rochev,Appelquist} the CJT effective action for
composite operators method has been extended to (Hermitian) quantum field theory models with fermions. As a 
result, a nonperturbative method has emerged for calculating various multi-fermion 
Green's functions based on functional equations of the Dyson-Schwinger type. Moreover, 
in this CJT effective action approach it is possible to 
investigate the possibility of dynamical generation of the fermion mass and chiral symmetry breaking, etc, as it was demonstrated, e.g., in the framework of the (1+1)-D GN model in Ref. 
\cite{Dorey}. And in the last case, i.e. in (1+1)-D, the results of the CJT effective action studies of the model are qualitatively the same as in the large-$N$ expansion technique. It is also worth mentioning  that the possibility of dynamically generating fermion mass in some non-Hermitian quantum field theory models has been investigated in Refs. \cite{Kanazawa2,Alexandre,Alexandre2,Mavromatos2}. Namely, in the first of these papers, the problem is considered within the framework of the $1/N$ expansion in the (3 + 1)-D NJL model (with a complex coupling constant), while in the remaining papers, for this purpose, the approach of the Dyson-Schwinger equation was used in non-Hermitian Yukawa-type models with additional four-fermion interaction term.

In the recent paper \cite{kkz} we have studied phase structure of the massless (2+1)-D GN model 
also using the CJT effective action method. It turns out that in this case, in contrast to (1+1)-D 
GN model, the CJT approach predicts a much richer phase structure compared to the result obtained
with other generally accepted nonperturbative methods, i.e. large-$N$ and optimized expansion 
techniques, etc. And each of the observed phases is associated with some dynamically generated 
(Hermitian) mass term of the Lagrangian. So in the present paper, we show that non-Hermiticity can 
arise spontaneously in the (2+1)-D GN model under consideration just in the framework of the CJT 
composite operator approach. It means that for a certain well-defined behavior of the bare coupling 
constant, the ground state of the system can be characterized by a dynamically arising non-Hermitian 
mass term of the Lagrangian, which can be both $\cP\cT$- and anti-$\cP\cT$ symmetric. 
 
The paper is organized as follows. Section \ref{IIA} presents the $N$-flavor massless 
(2+1)-dimensional Gross-Neveu model symmetric under several discrete transformations, two chiral 
transformations as well as with respect to spatial $\cP$ and time $\cT$ reflections. It also 
clarifies the question of how different fermion-antifermion structures (possible massive terms of 
the model Lagrangian) are transformed under the influence of $\cP\cT$.
In section \ref{IIB} the CJT effective action $\Gamma(S)$ of the composite bilocal and bifermion operator 
$\overline\psi (x)\psi (y)$ is constructed, which is actually the functional of the full 
fermionic propagator $S(x,y)$. In real situations, the propagator is a translation invariant 
solution of the stationary Schwinger-Dyson-type equation of the CJT effective action. 
In this section, the unrenormalized expression for $\Gamma(S)$ is obtained up to a first order 
in the bare coupling constant $G$. Based on this expression, we show in section III  that for a some
well-defined behavior of the coupling constant $G(\Lambda)$ vs $\Lambda$, there exist three different
renormalized, i.e. without ultraviolet divergences, solutions of the Schwinger-Dyson equation for 
the propagator. One of them corresponds to a phase in which a dynamically Hermitian mass 
term arises for fermions. The other two solutions correspond to two different phases with dynamically
emerging non-Hermitian mass terms. In each of these cases, the non-Hermiticity appears spontaneously 
in the originally Hermitian model, and it is accompanied by a real spectrum of fermions. 
Finally, in the section IV we show that the spontaneous 
non-Hermiticity of the model arises only in the chiral limit, i.e. if initially the Lagrangian of 
the model contains a (Hermitian) nonzero mass term, then non-Hermiticity does not arise.

\section{(2+1)-dimensional GN model and its CJT effective action}

\subsection{Model, its symmetries, etc}
\label{IIA}

We investigate the spontaneous (dynamical) generation of non-Hermitian mass terms in the 
simplest massless (2+1)-dimensional GN model. Its Lagrangian has the following form
\begin{eqnarray}
 L=\overline \psi_k\gamma^\nu i\partial_\nu \psi_k&+& \frac {G}{2N}\left
(\overline \psi_k\psi_k\right )^2,
\label{n1}
\end{eqnarray}
where for each $k=1,...,N$ the field $\psi_k\equiv \psi_k(t,x,y)$ is a (reducible) four-component
Dirac spinor (its spinor indices are omitted in Eq. (\ref{n1})), $\gamma^\nu$ 
($\nu=0,1,2$) are 4$\times$4 matrices acting in this four-dimensional spinor space (the algebra
of these $\gamma$-matrrices and their particular representation used in the present paper
is given in Appendix \ref{ApC}, where the matrices $\gamma^3,\gamma^5$ and $\tau=-i\gamma^3\gamma^5$
are also introduced), and the summation over repeated $k$- and $\nu$-indices is assumed in Eq. 
(\ref{n1}) and below. The bare coupling constant $G$ has a dimension of [mass]$^{-1}$. The 
Lagrangian is invariant under two discrete chiral transformations $\Gamma^5$ and $\Gamma^3$,
\begin{eqnarray}
 \Gamma^5:&~~&\psi_k(t,x,y)\to  \gamma^5\psi_k(t,x,y);~~
 \overline\psi_k(t,x,y)\to  -\overline\psi_k(t,x,y)\gamma^5,\nonumber\\
 \Gamma^3:&~~&\psi_k(t,x,y)\to  \gamma^3\psi_k(t,x,y);~~
 \overline\psi_k(t,x,y)\to  -\overline\psi_k(t,x,y)\gamma^3. 
 \label{n4}
\end{eqnarray}
Morerover, it is symmetric with respect to space parity $\cP$, time reversal $\cT$  and 
$\cP\cT$ symmetries, which we now discuss in more detail within the framework of model (\ref{n1}).

In (2+1) dimensions the space reflection, or parity, transformation $\cP$ is defined by
$(t,x,y)\stackrel{\mathcal P}{\longrightarrow} (t,-x,y)$. \footnote{In (2+1) spacetime dimensions, parity 
corresponds to inverting only one spatial axis \cite{Semenoff,Appelquist}, since the inversion 
of both axes is equivalent to rotating the entire space by $\pi$.} Moreover, we assume that an 
evident relation $\cP\cP={\bf 1}$ is valid. Let us derive the transformation of the spinor fields $\psi$ 
under $\cP$. To find this transformation, we postulate that the Lagrangian $L_0$ of the free massless 
spinor fields $\psi$ remains intact under space reflection $\cP$, i.e. $L_0$ equals to 
$\cP L_0 \cP$, where (below, for the sake of breavity we denote by $x$ and $x'$ the 
set of coordinates $(t,x,y)$ and $(t,-x,y)$, respectively) 
\begin{eqnarray}
L_0\equiv \overline\psi (x){\cal D}\psi(x), &&
\cP L_0 \cP=\overline{\psi^{\cP}} (x'){\cal D}'\psi^{\cP}(x'),~~~
{\cal D}=i\gamma^0\partial_0+i\gamma^1\partial_1+i\gamma^2\partial_2,\nonumber\\
{\cal D}'=\cP{\cal D}\cP&=&i\gamma^0\partial_0-i\gamma^1\partial_1+i\gamma^2\partial_2,~~
\psi^{\cP}(x')=\cP\psi (x)\cP,~~\overline{\psi^{\cP}} (x')=\cP\overline\psi (x)\cP.
\label{nn4}
\end{eqnarray}
It is not very difficult to notice from Eq. (\ref{nn4}) that $L_0$ is invariant under the 
action of $\cP$ only when 
\begin{eqnarray}
 \psi^{\cP}(x')\equiv\psi^{\cP}(t,-x,y)=  \gamma^5\gamma^1\psi(t,x,y);~{\rm i.e.}~~
 \overline{\psi^{\cP}} (x')\equiv\overline{\psi^{\cP}} (t,-x,y)= 
 \overline\psi(t,x,y)\gamma^5\gamma^1,
 \label{n2}
\end{eqnarray}
So the parity $\cP$ transformation of the fermion fields can be
defined by Eq. (\ref{n2}). Now, it is easy to show that the bifermion structure 
$\overline\psi(x)\psi(x)$ is $\cP$ invariant. Indeed, it is clear from Eqs. (\ref{nn4}) and 
(\ref{n2}) that
\begin{eqnarray}
\overline\psi(x)\psi(x)\stackrel{\mathcal P}{\longrightarrow}\cP\overline\psi (x)\cP
\cP\psi (x)\cP=\overline{\psi^{\cP}} (x')\psi^{\cP}(x')=\overline\psi(x)\gamma^5\gamma^1
 \gamma^5\gamma^1\psi(x)=\overline\psi(x)\psi(x).
\label{n3}
\end{eqnarray}
As a result, we conclude that the GN model (\ref{n1}) is $\cP$ invariant. In a similar way, one can 
find the $\cP$ transformations of some other Hermitian bispinor forms such as
\begin{eqnarray}
&&\overline\psi(x)i\gamma^5\psi(x)\stackrel{\mathcal P}{\longrightarrow}-\overline\psi(x)i\gamma^5
\psi(x),\nonumber\\
&&\overline\psi(x)i\gamma^3\gamma^5\psi(x)\stackrel{\mathcal P}{\longrightarrow}-
\overline\psi(x)i\gamma^3\gamma^5\psi(x),\nonumber\\
&&\overline\psi(x)i\gamma^3\psi(x)\stackrel{\mathcal P}{\longrightarrow}\overline\psi(x)i\gamma^3
\psi(x).
\label{nn3}
\end{eqnarray}
Now, let us consider the time reversal $\cT$ in the framework of the (2+1)-D GN model (\ref{n1}). 
In the (2+1)-dimensional spacetime it is defined as $(t,x,y)\stackrel{\mathcal T}{\longrightarrow} 
(-t,x,y)$, i.e. we suppose that $\cT\cT={\bf 1}$. To determine how the spinor fields $\psi$ are 
transformed under this operation in (2+1)-D
spacetime, we also assume from the very beginning (as in the case of spatial reflection $\cP$) that 
the Lagrangian $L_0$ (see in Eq. (\ref{nn4})) of free massless fermionic fields $\psi$ remains 
invariant with respect to $\cT$, i.e. $L_0=\cT L_0 \cT$, where (now, for the sake of breavity we 
denote below 
by $x$ and $x'$ the set of coordinates $(t,x,y)$ and $(-t,x,y)$, respectively) 
\begin{eqnarray}
\cT L_0 \cT=\overline{\psi^{\cT}} (x'){\cal D}'\psi^{\cT}(x'),~~~
\psi^{\cT}(x')=\cT\psi (x)\cT,~~\overline{\psi^{\cT}} (x')=\cT\overline\psi (x)\cT,
\label{n5}
\end{eqnarray}
and in this formula ${\cal D}'=\cT {\cal D}\cT$. In the following, it is very important to take 
into account that time-reversal operation $\cT$ (i) changes the sign of the time coordinate, $t\to -t$, 
and (ii) that it is an anti-linear or anti-unitary one, which means that its action on any complex 
number or matrix $C$ transforms it into the complex conjugate $C^*$, i.e. $\cT C \cT=C^*$ 
(for details, see, e.g., in Refs. \cite{Peskin2,Bender2,Chernodub}). Taking into account these 
(i) and (ii) properties of the $\cT$ transformation, we have
\begin{eqnarray}
{\cal D}'=i\gamma^{0*}\partial_0-i\gamma^{1*}\partial_1-i\gamma^{2*}\partial_2=
i\gamma^{0}\partial_0+i\gamma^{1}\partial_1-i\gamma^{2}\partial_2.
\label{n6}
\end{eqnarray}
In the last equality of Eq. (\ref{n6}) we used the relations $\gamma^{0*}=\gamma^{0}$, 
$\gamma^{1*}=-\gamma^{1}$ and $\gamma^{2*}=\gamma^{2}$ (see in Appendix \ref{ApC}). Now, it is 
rather evident from Eqs. (\ref{n5}) and (\ref{n6}) that $L_0$ is invariant under the 
action of $\cT$ only when 
\begin{eqnarray}
 \psi^{\cT}(x')\equiv\psi^{\cT}(-t,x,y)=  \gamma^5\gamma^2\psi(t,x,y);~{\rm i.e.}~~
 \overline{\psi^{\cT}} (x')\equiv\overline{\psi^{\cT}} (-t,x,y)= 
 \overline\psi(t,x,y)\gamma^5\gamma^2.
 \label{n7}
\end{eqnarray}
And just the relations (\ref{n7}) can be considered as a time reversal $\cT$ transformation of 
the spinor fields $\psi$. Now, taking into account Eqs. (\ref{n7}), it is possible to obtain the 
$\cT$ transformations of some Hermitian bispinor forms. In particular, 
\begin{eqnarray}
&&\overline\psi(x)\psi(x)\stackrel{\mathcal T}{\longrightarrow}\cT\overline\psi (x)\cT
\cT\psi (x)\cT=\overline{\psi^{\cT}} (x')\psi^{\cT}(x')=\overline\psi(x)\gamma^5\gamma^2
 \gamma^5\gamma^2\psi(x)=\overline\psi(x)\psi(x),\nonumber\\
&&~~~~~~~ \overline\psi(x)i\gamma^3\psi(x)\stackrel{\mathcal T}{\longrightarrow}\cT\overline\psi (x)\cT
\cT i\gamma^3\cT\cT\psi (x)\cT=\overline{\psi^{\cT}} (x')(-i)\gamma^3\psi^{\cT}(x')\nonumber\\
&&~~~~~~~~~~~~~~~~
=\overline\psi(x)\gamma^5\gamma^2(-i)\gamma^3 \gamma^5\gamma^2\psi(x)=-\overline\psi(x)i\gamma^3
\psi(x), \label{n8}
\end{eqnarray}
where we have used the anti-unitary property of the $\cT$ transformation, $\cT i\cT=-i$. In a 
similar way it is possible to find the $\cT$ transformations of some other Hermitian bispinor forms, 
\begin{eqnarray}
&&\overline\psi(x)i\gamma^5\psi(x)\stackrel{\mathcal T}{\longrightarrow}-\overline\psi(x)i\gamma^5
\psi(x),\nonumber\\
&&\overline\psi(x)i\gamma^3\gamma^5\psi(x)\stackrel{\mathcal T}{\longrightarrow}-
\overline\psi(x)i\gamma^3\gamma^5\psi(x).
\label{n9}
\end{eqnarray}
As it follows from Eqs. (\ref{n3}), (\ref{nn3}), (\ref{n8}) and (\ref{n9}), the Hermitian bispinor
structures $\overline\psi(x)\psi(x)$, $\overline\psi(x)i\gamma^5\psi(x)$ and 
$\overline\psi(x)i\gamma^3\gamma^5\psi(x)$ are $\cP\cT$ even (invariant), whereas the Hermitian 
$\overline\psi(x)i\gamma^3\psi(x)$ form is a $\cP\cT$ odd, i.e. it changes the sign under the action
of the $\cP\cT$ transformation, and due to this reason is called sometimes anti-$\cP\cT$-symmetric
\cite{Felski2}. 

From the point of 
view of further consideration, we are also interested in the behavior of such non-Hermitian 
structures as $\overline\psi(x)\gamma^5\psi(x)$ and $\overline\psi(x)\gamma^3\psi(x)$ with respect 
to $\cP$ and $\cT$ transformations. Obviously, under the action of $\cP$ they transform in the same 
way as the corresponding Hermitian forms in Eq. (\ref{nn3}). However, when reversing time, we have 
(below, we use evident relations $\cT\gamma^5\cT=\gamma^{5*}=-\gamma^5$ and $\cT\gamma^3\cT=\gamma^3$)
\begin{eqnarray}
&&\overline\psi(x)\gamma^5\psi(x)\stackrel{\mathcal T}{\longrightarrow}\overline\psi(x)\gamma^5
\psi(x),~~~
\overline\psi(x)\gamma^3\psi(x)\stackrel{\mathcal T}{\longrightarrow}
\overline\psi(x)\gamma^3\psi(x).
\label{n10}
\end{eqnarray}
Hence, $\overline\psi(x)\gamma^5\psi(x)$ is a $\cP\cT$-odd structure, but 
$\overline\psi(x)\gamma^3\psi(x)$ is a $\cP\cT$-even one. 

Below, we are going to study, using the CJT composite operator technique, the possibility 
for dynamical generation of the following mass terms in the model (\ref{n1})
\begin{eqnarray}
&&{\cal M}_H=i m_5\overline\psi(x)\gamma^5\psi(x)+i m_3\overline\psi(x)\gamma^3
\psi(x),\nonumber\\
&&{\cal M}_{NH1}=im_5\overline\psi(x)\gamma^5\psi(x)+ m_3\overline\psi(x)\gamma^3\psi(x),\nonumber\\
&&{\cal M}_{NH2}= m_5\overline\psi(x)\gamma^5\psi(x)+im_3\overline\psi(x)\gamma^3\psi(x)
\label{n13}
\end{eqnarray}
(note that in this formula and below it is assumed that $m_3$ and $m_5$ are real quantities). 
The mass term ${\cal M}_H$ is Hermitian and its dynamical appearance corresponds to spontaneous 
breaking of some of the above mentioned discrete symmetries of the model. Each of the remaining 
two mass terms in Eq. (\ref{n13}) is non-Hermitian; therefore, their dynamic appearance corresponds 
to the spontaneous non-Hermitian nature of the model (in addition to the spontaneous breaking of 
some of its discrete symmetries). Moreover, the mass term ${\cal M}_{NH1}$ is $\cP\cT$ even 
(symmetric), but the non-Hermitian mass term ${\cal M}_{NH2}$ 
is $\cP\cT$ odd, i.e. it changes the sign under this transformation. 

\subsection{CJT effective action of the model}
\label{IIB}

Before starting to consider the question of the spontaneous emergence of the  non-Hermiticity
of the model, it is necessary to add few words about the method for solving the problem, i.e. about 
the CJT composite operator approach.

Let us define $Z(K)$, the generating functional of the Green's functions of bilocal 
fermion-antifermion composite operators $\sum_{k=1}^N\overline\psi_k^\alpha(x)\psi_{k\beta}(y)$ 
in the framework of a (2+1)D GN model (\ref{n1}) (the corresponding technique 
for theories with four-fermion interaction is elaborated in details, e.g., in Ref. \cite{Rochev}) 
\begin{eqnarray}
 Z(K)\equiv\exp(iNW(K))=\int {\cal D}\overline\psi_k {\cal D}\psi_k \exp\Big
(i\Big [ I(\overline\psi,\psi)+\int d^3xd^3y\overline\psi_k^\alpha(x)K_\alpha^\beta(x,y)
\psi_{k\beta}(y) \Big ]\Big ),
 \label{36}
\end{eqnarray}
where $\alpha,\beta =1,2,3,4$ are spinor indices, $K_\alpha^\beta(x,y)$ is a bilocal source of 
the fermion bilinear composite field $\bar\psi_k^\alpha(x)\psi_{k\beta}(y)$ (recall that in 
all expressions the summation over repeated indices is assumed). 
\footnote{We denote a matrix element of an arbitrary matrix (operator) $\hat A$ acting in the 
four dimensional spinor space by the symbol $A^\alpha_\beta$, where the upper (low) index 
$\alpha $($\beta$) is the column (row) number of the matrix $\hat A$. In particular, the matrix 
elements of any $\gamma^\mu$ matrix is denoted by $(\gamma^\mu)^\alpha_\beta$. }
Moreover, $I(\bar\psi,\psi)=\int Ld^3x$, where $L$ is the Lagrangian (\ref{n1}) of a (2+1)-dimensional GN 
model under consideration. It is evident that
$$
I(\overline\psi ,\psi)=\int d^3xd^3y\overline\psi_k^\alpha(x)D_\alpha^\beta(x,y)\psi_{k\beta} (y)+
I_{int}(\overline\psi_k^\alpha\psi_{k\beta}),~~D_\alpha^\beta(x,y)=
\left(\gamma^\nu\right)_\alpha^\beta i\partial_\nu\delta^3(x-y),
$$\vspace{-0.5cm}
\begin{eqnarray}
I_{int}&=& \frac {G}{2N}\int d^3x\left (\bar \psi_k\psi_k\right )^2
=\frac {G}{2N}\int d^3xd^3td^3ud^3v\delta^3 (x-t)\delta^3 (t-u)\delta^3 (u-v) \overline
\psi_k^\alpha(x)\delta_\alpha^\beta\psi_{k\beta}(t) \overline\psi_l^\rho(u)
\delta_\rho^\xi\psi_{l\xi}(v).
 \label{360}
\end{eqnarray}
Note that in Eq. (\ref{360}) and similar expressions below, $\delta^3(x-y)$ denotes the 
three-dimensional Dirac delta function. There is an alternative expression for $Z(K)=\exp(iNW(K))$,
\begin{eqnarray}
\exp(iNW(K))&=&\exp\Big (iI_{int}\Big (-i\frac{\delta}{\delta K}\Big )\Big )\int 
{\cal D}\overline\psi_k {\cal D}\psi_k \exp\Big (i
\int d^3xd^3y\overline\psi_k(x)\Big [D(x,y)+K(x,y)\Big ]\psi_k (y)\Big )
\nonumber\\&=&\exp\Big (iI_{int}\Big (-i\frac{\delta}{\delta K}\Big )\Big )\Big 
[\det\big (D(x,y)+K(x,y)\big )\Big ]^N\nonumber\\&=&\exp\Big (iI_{int}\Big 
(-i\frac{\delta}{\delta K}\Big )\Big )\exp \Big [N{\rm Tr}\ln \big (D(x,y)+K(x,y)\big )\Big ],
\label{036}
\end{eqnarray}
where instead of each bilinear form $\bar\psi_k^\alpha(s)\psi_{k\beta}(t)$ appearing in $I_{int}$ of
the Eq. (\ref{360}) we use a variational derivative $-i\delta /\delta K^\beta_\alpha (s,t)$.
Moreover, the Tr-operation in Eq. (\ref{036}) means the trace both over spacetime and spinor coordinates. 
The effective action (or CJT effective action) of the composite bilocal and bispinor operator 
$\bar\psi_k^\alpha(x)\psi_{k\beta}(y)$ is defined as a functional $\Gamma (S)$ of the full 
fermion propagator $S^\alpha_\beta(x,y)$ by a Legendre transformation of the functional 
$W(K)$ entering in Eqs. (\ref{36}) and (\ref{036}),
\begin{eqnarray}
\Gamma (S)=W(K)-\int d^3xd^3y S^\alpha_\beta(x,y)K_\alpha^\beta(y,x),
\label{0360}
\end{eqnarray}
where
\begin{eqnarray}
S^\alpha_\beta(x,y)=\frac{\delta W(K)}{\delta K_\alpha^\beta(y,x)}.
 \label{37}
\end{eqnarray}
Taking into account the relation (\ref{36}), it is clear that $S(x,y)$ is the full fermion 
propagator at $K(x,y)=0$. Hence, in order to construct the CJT effective action $\Gamma (S)$ of Eq. 
(\ref{0360}), it is necessary to solve Eq. (\ref{37}) with respect to $K$ and then to use the 
obtained expression for $K$ (it is a functional of $S$) in Eq. (\ref{0360}). It is clear from the 
definition (\ref{0360})-(\ref{37}) that
\begin{eqnarray}
\frac{\delta\Gamma (S)}{\delta S^\alpha_\beta(x,y)}=
\int d^3ud^3v\frac{\delta W(K)}{\delta K^\mu_\nu(u,v)}\frac{\delta K^\mu_\nu(u,v)}
{\delta S^\alpha_\beta(x,y)}-K_\alpha^\beta(y,x)-\int d^3ud^3v S_\mu^\nu(v,u)
\frac{\delta K^\mu_\nu(u,v)}{\delta S^\alpha_\beta(x,y)}.
 \label{037}
\end{eqnarray}
(In Eq. (\ref{037}) and below, the Greek letters $\alpha,\beta,\mu,\nu,$ etc, also denote
the spinor indices, i.e. $\alpha,...\nu,...=1,...,4$.) Now, due to the relation (\ref{37}), it is 
easy to see that the first term in Eq. (\ref{037}) 
cansels there the last term, so
\begin{eqnarray}
\frac{\delta\Gamma (S)}{\delta S^\alpha_\beta(x,y)}=-K_\alpha^\beta(y,x).
\label{370}
\end{eqnarray}
Hence, in the true GN theory, in which bilocal sources $K_\alpha^\beta(y,x)$ are zero, the full 
fermion propagator is a solution of the following stationary equation,
\begin{eqnarray}
\frac{\delta\Gamma (S)}{\delta S^\alpha_\beta(x,y)}=0.
\label{0370}
\end{eqnarray}
Note that in the nonperturbative CJT approach the stationary/gap equation (\ref{0370}) 
for fermion propagator $S^\beta_\alpha(x,y)$  is indeed a 
Schwinger--Dyson equation \cite{Rochev}.
Further, in order to simplify the calculations and obtain specific information about 
the phase structure of the model, we calculate the effective action (\ref{0360}) up to a 
first order in the coupling $G$. In this case (the detailed calculations are given in Appendix B
of Ref. \cite{kkz})
\begin{eqnarray}
\Gamma (S)&=&-i{\rm Tr}\ln \big (-iS^{-1}\big )+\int d^3xd^3y S^\alpha_\beta(x,y)D_\alpha^\beta(y,x)
\nonumber\\
&+&\frac{G}2\int d^3x \Big [{\rm tr}S(x,x)\Big ]^2 -\frac{G}{2N}\int d^3x~ {\rm tr}\Big 
[S(x,x)S(x,x)\Big ].
 \label{n420}
\end{eqnarray}
Notice that in Eq. (\ref{n420}) the symbol tr means the trace of an operator 
over spinor indices only, but Tr is the trace operation both over spacetime coordinates and spinor 
indices. Moreover, there the operator $D(x,y)$ is introduced in Eq. (\ref{360}). The stationary 
equation (\ref{0370}) for the CJT effective action (\ref{n420}) looks like 
\begin{eqnarray}
0&=&i\Big [S^{-1}\Big ]^\beta_\alpha(x,y)+D_\alpha^\beta(x,y)+G\delta^\beta_\alpha\delta^3
(x-y)~{\rm tr}S(x,y)-\frac GN S^\beta_\alpha(x,y)\delta^3 (x-y).
\label{n0420}
\end{eqnarray}
Now suppose that $S(x,y)$ is a translationary invariant operator. Then 
\begin{eqnarray}
S^\beta_\alpha(x,y)\equiv S^\beta_\alpha(z)&=&\int\frac{d^3p}{(2\pi)^3}
\overline{S^\beta_\alpha}(p)e^{-ipz},~~~\overline{S^\beta_\alpha}(p)=\int d^3z S^\beta_\alpha(z)e^{ipz},\nonumber\\
\Big (S^{-1}\Big )^\beta_\alpha(x,y)&\equiv& \Big (S^{-1}\Big )^\beta_\alpha(z)=\int\frac{d^3p}{(2\pi)^3}
\overline{(S^{-1})^\beta_\alpha}(p)e^{-ipz},
\label{n43}
\end{eqnarray}
where $z=x-y$ and $\overline{S^\beta_\alpha}(p)$ is a Fourier transformation of $ S^\beta_\alpha(z)$.
After Fourier transformation, the Eq. (\ref{n0420}) takes the form
\begin{eqnarray}
\overline{(S^{-1})^\beta_\alpha}(p)-i p_\nu(\gamma^\nu)^\beta_\alpha=iG\delta^\beta_\alpha\int\frac{d^3q}{(2\pi)^3}~{\rm tr}\overline{S}(q)
-i\frac GN \int\frac{d^3q}{(2\pi)^3}\overline{S^\beta_\alpha}(q).
 \label{n043}
\end{eqnarray}
It is clear from Eq. (\ref{n043}) that in the framework of the four-fermion model (\ref{n1}) 
the Schwinger-Dyson equation for fermion propagator $\overline{S}(p)$ reads in the first order
in $G$ like the Hartree-Fock equation for its self-energy operator $\Sigma(p)$. In particular, 
the first and second terms on the right-hand side of Eq. (\ref{n043}) are, respectively, the 
so-called Hartree and Fock contributions to the fermion self energy (for details, see, e.g., 
the section 4.3.1 in Ref. \cite{Buballa}). 

Finally note that both the CJT effective action (\ref{n420}) and its stationary equation 
(\ref{n0420})-(\ref{n043}), in which $G$ is a bare coupling constant, contain ultraviolet (UV)
divergences and need to be renormalized. In the next sections, we will find out at what behavior 
of the bare coupling constant $G\equiv G(\Lambda)$ vs $\Lambda$ it is possible to renormalize 
the stationarity equation (\ref{n043}), the finite solution of which corresponds 
at $\Lambda\to\infty$ to the dynamical appearance of the mass terms of the form (\ref{n13}) 
in the Lagrangian. 

\section{Possibility for dynamical generation  of the mass terms (\ref{n13})}
\subsection{Dynamical generation of the Hermitiam mass ${\cal M}_H$}

Let us explore the possibility that the solution of the gap equation (\ref{n043}) has the form 
\begin{eqnarray}
\overline{S^{-1}}(p)=i(\hat p+i\gamma^5 m_5+i\gamma^3 m_3), ~~~{\rm i.e.}~~~
\overline {S}(p)=-i\frac{\hat p+i\gamma^5 m_5+i\gamma^3 m_3}{p^2-(m_3^2+m_5^2)}.
 \label{0471}
\end{eqnarray}
It corresponds to a dynamically generated mass term of the form ${\cal M}_H=\left (m_5\overline\psi i\gamma^5\psi
+m_3\overline\psi i\gamma^3\psi\right )$ in the Lagrangian (\ref{n1}) (the Hermitian matrices $\gamma^{3,5}$
are presented in Appendix \ref{ApC}). Since we suppose that $m_5$ and $m_3$ are some real 
numbers, this mass term is a Hermitian one. And it is not invariant under each of the 
discrete transformations (\ref{n4}) or (\ref{nn4}) (at nonzero $m_3$ and $m_5$).
Substituting Eq. (\ref{0471}) into Eq. (\ref{n043}), 
one can obtain for $m_3$ and $m_5$ the following system of gap equations
\begin{eqnarray}
m_3&=&\frac {im_3G}{N}\int\frac{d^3p}{(2\pi)^3}\frac{1}{p^2-(m_3^2+m_5^2)},\nonumber\\
 m_5&=&\frac {im_5G}{N}\int\frac{d^3p}{(2\pi)^3}\frac{1}{p^2-(m_3^2+m_5^2)}.\label{49}
\end{eqnarray}
After a Wick rotation in Eq. (\ref{49})
to Euclidean energy-momentum, i.e. $p_0\to i p_0$, we see that $(m_3,m_5)$ should obey the 
equation system (in which $p^2=p_0^2+p_1^2+p^2_2$)
\begin{eqnarray}
\frac{m_3}{G}&=&\frac {m_3}{N}\int\frac{d^3p}{(2\pi)^3}\frac{1}{p^2+(m_3^2+m_5^2)},\nonumber\\
\frac{m_5}{G}&=&\frac {m_5}{N}\int\frac{d^3p}{(2\pi)^3}\frac{1}{p^2+(m_3^2+m_5^2)}.
\label{049}
\end{eqnarray}
This system contains UV-divergent integrals, i.e. it is unrenormalized. For its 
regularization, we use the spherical coordinate system when 
$\int d^3pf(\sqrt{p_0^2+p_1^2+p_2^2})=4\pi\int_0^\infty p^2dpf(p)$ and 
$p=\sqrt{p_0^2+p_1^2+p_2^2}$. Then, cutting the region of 
integration in the obtained one-dimensional UV-divergent 
integral by $\Lambda$, we have for $(m_3,m_5)$ the {\it regularized} gap equations
\begin{eqnarray}
\frac {m_3}G&=&\frac{m_3}{2N\pi^2}\int_0^\Lambda\frac{p^2}{p^2+m_3^2+m_5^2}dp,\nonumber\\
 \frac {m_5}G&=&\frac{m_5}{2N\pi^2}\int_0^\Lambda\frac{p^2}{p^2+m_3^2+m_5^2}dp.
 \label{490}
\end{eqnarray}
Notice that at $\Lambda\to\infty$ an integral term in Eqs.  (\ref{490}) has the 
following asymptotic expansion
\begin{eqnarray}
\int_0^\Lambda\frac{p^2}{p^2+m_3^2+m_5^2}dp=\Lambda-\frac \pi 2\sqrt{m_3^2+m_5^2}+
\sqrt{m_3^2+m_5^2}{\cal O}\left (\frac {\sqrt{m_3^2+m_5^2}}\Lambda\right ).
 \label{4900}
\end{eqnarray}
Hence, taking into account the expansion (\ref{4900}), the UV divergence can be removed from 
the gap equations (\ref{490}) if we require 
 (it is clear from the form of this equation system) the following behavior of the bare coupling 
 constant $G\equiv G(\Lambda)$ vs $\Lambda$,
\begin{eqnarray}
\frac 1{G(\Lambda)}=\frac{1}{2N\pi^2}\Big (\Lambda+g\frac \pi 2+g{\cal O}\Big (\frac {g}
\Lambda\Big )\Big ),
\label{0490}
\end{eqnarray}
where $g$ is a finite $\Lambda$-independent and renormalization group invariant quantity, and 
it can also be considered as a new free parameter of the model. Now, comparing Eqs. 
(\ref{0490})
and (\ref{490}), we obtain in the limit $\Lambda\to\infty$ the 
following renormalized, i.e. without UV divergences, gap equations for the masses $m_3$ and $m_5$
\begin{eqnarray}
m_3\left (g+\sqrt{m_3^2+m_5^2}\right )&=&0,\nonumber\\
m_5\left (g+\sqrt{m_3^2+m_5^2}\right )&=&0.
\label{40400}
\end{eqnarray}
Hence, at $g>0$ only a trivial solution of the gap equations (\ref{40400}) exists, $m_3=m_5=0$,
and all discrete symmetries of the model remain intact. 
However, at $g<0$ there are two solutions, (i) $m_3=0, m_5=0$ and (ii)
$m_3=|g|\cos\alpha, m_5=|g|\sin\alpha$ (where $0\le \alpha\le \pi /2$ is some arbitrary fixed 
angle), of the system (\ref{40400}) of gap equations. 
To find which of the solutions of the gap equations, (i) or (ii), is more preferable in this 
case, it is necessary to consider the so-called CJT effective potential $V(S)$ of the model, 
which is defined on the basis of the CJT effective action (\ref{n420}) by the following 
relation \cite{CJT,Casalbuoni}
\begin{eqnarray}
V(S)\int d^3x\equiv -\Gamma(S)\Big |_{\rm transl.-inv.~S(x,y) },
 \label{n470}
\end{eqnarray}
where $S(x,y)$ is a translation invariant quantity, i.e. $S(x,y)\equiv S(x-y)$, as assumed
in Eq. (\ref{n43}). It is evident that for arbitrary values of the bare coupling
constant $G$ the CJT effective potential (\ref{n470}) is UV-divergent unrenormalized quantity.
However, if $G$ is constrained by the condition (\ref{0490}), then the UV divergences of $V(S)$ are 
eliminated, and for fermion propagator of the form (\ref{0471}) it looks at $\Lambda\to\infty$ like  
(up to unessential $m_3$- and $m_5$-independent infinite constant)
\begin{eqnarray}
V(S)\equiv V_H(m_3,m_5)=\frac{1}{6\pi}\left (2(m_3^2+m_5^2)^{3/2}+3g (m_3^2+m_5^2)\right ).
 \label{492}
\end{eqnarray} 
It is clear from Eq. (\ref{492}) that at $g<0$ the effective potential $V_H$ takes on the solution 
(ii) the value $-|g|^3/(6\pi)$, and this quantity is smaller than $V_H(m_3=0,m_5=0)=0$. This allows 
us to conclude that if in the original model (\ref{n1}) the bare coupling constant $G$ behaves vs 
$\Lambda$ like in the expression (\ref{0490}) with $g<0$, then the system undergoes dynamic 
generation of the $m_3=|g|\cos\alpha$ and $m_5=|g|\sin\alpha$ masses, i.e. a phase with spontaneous 
violation of all discrete symmetries is realized in the model (if $\alpha\ne 0,\pi/2$). But if 
$\alpha =0$ then only $\cP\cT$-odd $m_3=|g|$ mass term is generated, and $\Gamma^3$ chiral symmetry 
(\ref{n4}) is dynamically violated. However, at $\alpha =\pi/2$ only $\cP\cT$-even $m_5=|g|$ mass 
term appears dynamically, and in this case chiral $\Gamma^5$ symmetry is broken spontaneously. 
Finally notice that at $g<0$ in all above mentioned cases, i.e. at arbitrary values of the angle 
parameter $\alpha$, the genuine physical fermion mass, which is indeed a pole of the 
fermion propagator (\ref{0471}), is equal to $M_F=\sqrt{m_3^2+m_5^2}\equiv |g|$.

In terms of dimensionless bare coupling constant $\lambda\equiv\lambda(\Lambda)=
\Lambda G(\Lambda)$, where $G(\Lambda)$ is given by Eq. (\ref{0490}), the situation looks as 
follows. It is clear that in this case for a sufficiently 
high values of $\Lambda\gg |g|$ both the dimensional bare coupling $G (\Lambda)$ and
the dimensionless coupling $\lambda$ are positive. In addition, it is easy to see
that at $\Lambda\to\infty$ the dimensionless bare coupling $\lambda$ tends to
the quantity $\lambda_{crit}\equiv 2N\pi^2$, which in fact is the UV-stable fixed point of the model
(see in Ref. \cite{kkz}).
Then the relation 
\begin{eqnarray}
\lambda -\lambda_{crit} \sim -\frac{2\pi^2Ng}{\Lambda}
 \label{495}
\end{eqnarray} 
can be obtained. It follows from Eq. (\ref{495}) that on the positive $\lambda$-semiaxis the 
UV-fixed point $\lambda_{crit}$ separates the symmetric phase from the one in which fermions 
are massive.  Indeed, if $\lambda>\lambda_{crit}$ then, as it is clear from Eq. 
(\ref{495}), the parameter $g$ must be negative. In this case dynamical generation of the Hermitian 
mass term ${\cal M}_H=\left (m_5\overline\psi i\gamma^5\psi +m_3\overline\psi i\gamma^3\psi\right )$
occurs in the Lagrangian (which indeed corresponds to a physical mass $M_F$ of fermion quasiparticles 
equal to $|g|$). In contrast, at $\lambda<\lambda_{crit}$ we have $g>0$ from Eq. (\ref{495}) and 
symmetric phase of the model (see the text below Eq. (\ref{40400})). \label{IIIA}

\subsection{Dynamical generation of the non-Hermitiam mass terms ${\cal M}_{NH1}$ and 
${\cal M}_{NH2}$}

\underline{Generation of the ${\cal M}_{NH1}$ mass term.}  First, let us explore, using the CJT 
approach, the possibility of the dynamic appearance of a 
non-Hermitian and $\cP\cT$ symmetric mass term ${\cal M}_{NH1}$ (\ref{n13}) in the original (2+1)-D 
GN model (\ref{n1}). In this case we should find the solution of the gap equation (\ref{n043}) in 
the form
\begin{eqnarray}
\overline{S^{-1}}(p)=i(\hat p+i\gamma^5 m_5+\gamma^3 m_3), ~~~{\rm i.e.}~~~
\overline {S}(p)=-i\frac{\hat p+i\gamma^5 m_5+\gamma^3 m_3}{p^2-(m_5^2-m_3^2)},
 \label{0472}
\end{eqnarray}
where $m_3$ and $m_5$ are real quantities. In addition, we suppose that $m_5^2\ge m_3^2$. 
Substituting Eq. (\ref{0472}) into the CJT stationary equation (\ref{n043}), one can obtain for 
$m_3$ and $m_5$ the 
UV-divergent system of gap equations. In it, one can also perform a Wick rotation and then to 
integrate over spherical angles. Finally, after cutting off the region of integration by $\Lambda$, 
we obtain for $m_3$ and $m_5$ a {\it regularized} system, which looks like
\begin{eqnarray}
\frac {m_3}G&=&\frac{m_3}{2N\pi^2}\int_0^\Lambda\frac{p^2}{p^2+m_5^2-m_3^2}dp,\nonumber\\
 \frac {m_5}G&=&\frac{m_5}{2N\pi^2}\int_0^\Lambda\frac{p^2}{p^2+m_5^2-m_3^2}dp.
 \label{491}
\end{eqnarray}
Since in our consideration it is assumed that $m_5^2\ge m_3^2$, one can apply in Eq. (\ref{491})
the asymptotic expansion (\ref{4900}) and find that the bare coupling constant $G$, which behaves
vs $\Lambda$ like in Eq. (\ref{0490}), removes at $\Lambda\to\infty$ all UV-divergences from the 
gap system (\ref{491}). And its finite, i.e. {\it renormalized}, form reads 
\begin{eqnarray}
m_3\left (g+\sqrt{m_5^2-m_3^2}\right )&=&0,\nonumber\\
m_5\left (g+\sqrt{m_5^2-m_3^2}\right )&=&0.
\label{40401}
\end{eqnarray}
There are several solutions of the system (\ref{40401}). To find the more preferable from the energy 
point of view, we again use the CJT effective potential (\ref{n470}), which, after substituting the 
expressions (\ref{0472}) and (\ref{0490}) into Eq. (\ref{n420}), takes the form (recall that in our 
case $m_5^2-m_3^2\ge 0$)
\begin{eqnarray}
V(S)\equiv V_{NH1}(m_3,m_5)=\frac{1}{6\pi}\left (2(m_5^2-m_3^2)^{3/2}+3g (m_5^2-m_3^2)\right ).
 \label{493}
\end{eqnarray} 
Hence, at $g>0$ its global minimum lies at the point $m_5=m_3=0$, and dynamical mass generation is 
absent. But at $g<0$ the global minimum of the $V_{NH1}(m_3,m_5)$ is $-|g|^3/(6\pi)$, and it is 
achieved at arbitrary $(m_3,m_5)$ point such that $m_5^2-m_3^2=g^2$, i.e. when $m_3=|g|\sinh\beta$ 
and $m_5=|g|\cosh\beta$, where $\beta\in \mathbb R$. Note that such a structure of the global 
minimum point of the model appears due to the emergent symmetry of the CJT effective potential 
(\ref{493}) with respect to non-Unitary transformations
\begin{eqnarray}
\left (\begin{array}{cc}
m_5\\
m_3
\end{array}\right )\to 
\left (\begin{array}{cc}
\cosh\beta & \sinh\beta\\
 \sinh\beta &\cosh\beta
\end{array}\right )\left (\begin{array}{cc}
m_5\\
m_3
\end{array}\right ).
\label{4930}
\end{eqnarray}
The energies of all these ground states at which 
a non-Hermitian and $\cP\cT$-symmetric fermion mass term ${\cal M}_{NH1}$ (\ref{n13}) arises 
spontaneously in the system are equal to each other and, moreover, coincide with the energy of any 
vacuum state corresponding to the dynamic appearance of a Hermitian mass term ${\cal M}_{H}$ of 
fermions in the system (see in the previous subsection). The singularity of the fermion propagator 
(\ref{0472}) corresponds to the fact that the quasiparticle excitations of each of these 
non-Hermitian and $\cP\cT$-even ground states of the system  have a real mass spectrum, i.e. 
their masses are real and also equal to the same value
$M_F=\sqrt{m_5^2-m_3^2}\equiv |g|$, which is observed in the case with Hermitian vacuum (see in the 
subsection \ref{IIIA}). 

\underline{Generation of the ${\cal M}_{NH2}$ mass term.} Omitting unnecessary details, it can be 
shown in exactly the same way that for the same dependence 
(\ref{0490}) of the bare coupling constant $G$ vs $\Lambda$, there exists a nontrivial solution of 
the renormalized stationary (Dyson-Schwinger) equation (\ref{n043}) of the form 
\begin{eqnarray}
\overline{S^{-1}}(p)=i(\hat p+\gamma^5 m_5+i\gamma^3 m_3), ~~~{\rm i.e.}~~~
\overline {S}(p)=-i\frac{\hat p+\gamma^5 m_5+i\gamma^3 m_3}{p^2-(m_3^2-m_5^2)},
 \label{0473}
\end{eqnarray}
which corresponds to spontaneous generation at $g<0$ of the non-Hermitian but $\cP\cT$-odd mass 
term ${\cal M}_{NH2}$ (\ref{n13}) in the model. In this case $m_3=|g|\cosh\omega$ 
and $m_5=|g|\sinh\omega$, where $\omega\in \mathbb R$. And in this phase fermion propagator 
(\ref{0473}) describes, for any real value of $\omega$, the quasiparticles with the same real mass 
$M_F=\sqrt{m_3^2-m_5^2}\equiv |g|$.

\underline{Conclusions.} As a result, we see that at $\lambda>\lambda_{crit}=2N\pi^2$, where 
$\lambda\equiv 
\Lambda G(\Lambda)$ is the dimensionless coupling constant of the model (see in the last paragraph 
of the previous subsection \ref{IIIA}), there might appear, on the same footing, three different 
phases in the framework of the Hermitian massless (2+1)-D GN model (\ref{n1}). One of them is 
characterized by dynamical appearance of the Hermitian mass term ${\cal M}_{H}$ in the Lagrangian, 
i.e. the ground state of the system remains Hermitian. However, 
in each of the remaining two phases, a non-Hermitian mass term, ${\cal M}_{NH1}$ or ${\cal M}_{NH2}$,
is dynamically generated in the Lagrangian of the model. And in this case non-Hermiticity of the 
model appears spontaneously, which is accompanied by a real mass spectrum of quasiparticle 
excitations.  In the first case, when ${\cal M}_{NH1}$ is 
generated, the non-Hermitian ground state of the model remains $\cP\cT$ symmetrical, but when 
${\cal M}_{NH2}$ appears dynamically, the $\cP\cT$-invariance of the model is broken spontaneously.
Moreover, we note that the ground states of all these three different phases, both Hermitian and 
non-Hermitian, have the same energy density equal to $-|g|^3/(6\pi)$, i.e. they are degenerate.
The similar result was obtained in Ref. \cite{Chernodub2} within (3+1)-D NJL model, where it was 
shown that in the chiral limit the uniform non-Hermitian ground state has the same (finite) free
energy density as the usual Hermitian ground state.

Finally, two remarks. (i) The phenomenon of spontaneous emergence of non-Hermiticity
in the massless (2+1)-D GN model (\ref{n1}) is characteristic only for finite values of $N$ and 
cannot be observed in the framework of the large-$N$ expansion technique. Indeed, it might occur 
only at $\lambda>\lambda_{crit}=2N\pi^2$. So at $N\to\infty$ it is absent. (ii) In the (3+1)-D NJL 
model, as it is proved in Ref. \cite{Chernodub2}, the spontaneous non-Hermiticity occurs only in the 
chiral limit, i.e. if bare quark mass is zero. However, if it is a nonzero quantity, then this 
effect does not appear. In the next section, we show that in the framework of the
(2+1)-D GN model (\ref{n1}) with nonzero (Hermitian) bare mass of fermions, non-Hermiticity also 
cannot arise spontaneously.\label{IIIB}

\section{The case of nonzero Hermitian bare $m_5$ mass}

Let us study the dynamical symmetry breaking in the (2+1)-D GN model (\ref{n1}) when its Lagrangian contains, e.g., a nonzero bare Hermitian chiral mass $m_5$, i.e. the Lagrangian of the model looks like
\begin{eqnarray}
 L=\overline \psi_k\gamma^\nu i\partial_\nu \psi_k+ i m_{05}\overline \psi_k\gamma^5\psi_k+
 \frac {G}{2N}\left (\overline \psi_k\psi_k\right )^2.
\label{m01}
\end{eqnarray}
In this case it is invariant only under discrete chiral $\Gamma^3$ and $\cP\cT$ transformations 
(other discrete symmetries $\Gamma^5$, $\cP$ and $\cT$ of the massless model (\ref{n1})
are violated explicitly, as it is shown in section \ref{IIA}, by the mass term of the Lagrangian 
(\ref{m01})). In the present section, we are going to consider in the framework of the model 
(\ref{m01}) the possibility of dynamical generation in it of the $m_3$-mass term of the form 
$\kappa m_{3}\overline \psi\gamma^3\psi$, where $\kappa$ is equal to $1$  or $i$ and $m_3$ is real. If $\kappa =i$, 
then both the mass term and the model remain Hermitian, however if $\kappa =1$ then 
the ground state of the model corresponds to spontaneous emergence of non-Hermiticity in it. 
Note that in both cases the discrete $\Gamma^3$ symmetry (\ref{n4}) of the model is spontaneously 
broken down.

The consideration is again performed on the basis of the CJT effective action (\ref{n420}) and its
stationary equation (\ref{n0420}), in which this time $D(x,y)=[\gamma^\nu i\partial_\nu+
i m_{05}\gamma^5]\delta^3(x-y)$, i.e. $\overline D(p)=\hat p+i m_{05}\gamma^5$.
Fourier transformation of this gap equation reads as 
\begin{eqnarray}
-i\overline{(S^{-1})^\beta_\alpha}(p)=p_\nu(\gamma^\nu)^\beta_\alpha+i m_{05}\gamma^5+
G\delta^\beta_\alpha\int\frac{d^3q}{(2\pi)^3}~{\rm tr}\overline{S}(q)-
\frac GN \int\frac{d^3q}{(2\pi)^3}\overline{S^\beta_\alpha}(q).
 \label{m02}
\end{eqnarray}
Here we explore the possibility that the solution of this gap equation has the form 
\begin{eqnarray}
\overline{S^{-1}}(p)=i(\hat p+i\gamma^5 m_5+\kappa\gamma^3 m_3), ~~~{\rm i.e.}~~~
\overline {S}(p)=-i\frac{\hat p+i\gamma^5 m_5+\kappa\gamma^3 m_3}{p^2-(m_5^2-\kappa^2 m_3^2)}.
 \label{m03}
\end{eqnarray}
It corresponds to a dynamically generated mass term of the form 
$\kappa m_3\overline\psi \gamma^3\psi$ in the Lagrangian (\ref{m01}) (the Hermitian matrices
$\gamma^{3,5}$ are presented in Appendix \ref{ApC}). Moreover, we suppose that for each value of 
$\kappa =1,i$ the mass parameters $m_5$ and $m_3$ both in Eq. (\ref{m03}) and throughout the 
consideration are some real numbers.
Substituting Eq. (\ref{m03}) into Eq. (\ref{m02}) and taking into account the technical 
details discussed in previous two sections, one can obtain for $m_3$ and $m_5$ 
the following system of gap equations (in which $p^2=p_0^2-p_1^2-p^2_2$)
\begin{eqnarray}
m_3&=&\frac {im_3G}{N}\int\frac{d^3p}{(2\pi)^3}\frac{1}{p^2-(m_5^2-\kappa^2 m_3^2)},\nonumber\\
 m_5&=&m_{05}+\frac {im_5G}{N}\int\frac{d^3p}{(2\pi)^3}\frac{1}{p^2-(m_5^2-\kappa^2 m_3^2)}.
\label{m04}
\end{eqnarray}
 After a Wick rotation in Eq. (\ref{m04})
to Euclidean energy-momentum, i.e. $p_0\to i p_0$, it is possible to use there a spherical coordinate 
system. Then we integrate in the obtained expressions over spherical angles and cut off by $\Lambda$ 
the resulting one-dimensional integral. As a result, we see that the set $(m_3,m_5)$ should obey the
following {\it regularized} system 
\begin{eqnarray}
\frac{m_3}{G}&=&\frac {m_3}{N}\int\frac{d^3p}{(2\pi)^3}\frac{1}{p^2+(m_5^2-\kappa^2 m_3^2)}=
\frac {m_3}{2N\pi^2}\int_0^\Lambda p^2dp\frac{1}{p^2+(m_5^2-\kappa^2 m_3^2)},\nonumber\\
\frac{m_5}{G}-\frac{m_{05}}{G}&=&
\frac {m_5}{N}\int\frac{d^3p}{(2\pi)^3}\frac{1}{p^2+(m_5^2-\kappa^2 m_3^2)}=
\frac {m_5}{2N\pi^2}\int_0^\Lambda p^2dp\frac{1}{p^2+(m_5^2-\kappa^2 m_3^2)}.\label{m05}
\end{eqnarray}
Taking into account in these {\it regularized} equations the asymptotic expansion (\ref{4900}),
we see that the system of equations (\ref{m05}) can be {\it renormalized} if we demand the same
behavior (\ref{0490}) of the bare coupling constant $G$ vs $\Lambda$, as well as that 
$\frac{m_{05}}{G}=\pm\frac{m^2}{4\pi N}$, in addition (here $m\ne 0$). Then $(m_3,m_5)$ should obey 
the following finite gap system 
\begin{eqnarray}
m_3\left (g+\sqrt{m_5^2-\kappa^2 m_3^2}\right )&=&0,\nonumber\\
m_5\left (g+\sqrt{m_5^2-\kappa^2 m_3^2}\right )&=&\pm m^2.
\label{m06}
\end{eqnarray}
Recall that both parameters $g$ and $m$ are some finite and renormalization group invariant
quantities with dimension of [mass]. Since $m\ne 0$ (otherwise the bare mass $m_{05}$ would be equal 
to zero), the expression in parentheses of Eq. (\ref{m06}) is always nonzero. So the solution of Eq.
(\ref{m06}) is such that $m_3=0$, and $m_5$ should obey the equation 
\begin{eqnarray}
m_5\left (g+|m_5|\right )&=&\pm m^2.
\label{m07}
\end{eqnarray}
Hence, we conclude that both for $\kappa=i$ and $\kappa=1$ there no exist solutions of the gap 
equations (\ref{m06}) with $m_3\ne 0$. It means that in the case when bare mass $m_{05}\ne 0$, 
neither Hermitian $ im_{3}\overline \psi\gamma^3\psi$ nor non-Hermitian $m_{3}\overline \psi\gamma^3\psi$ 
mass terms can arise dynamically in the model. In a similar way, it can be shown that if some other
Hermitian bare mass term (instead of the chiral $m_{05}$ considered in this section) is present in 
the Lagrangian (\ref{n1}) (for example, the Dirac $ \overline \psi\psi$, the Haldane 
$ i\overline \psi\gamma^3\gamma^5\psi$-mass terms, etc.), then it is this mass term that 
is modified in the framework of the CJT composite approach. The dynamic emergence of other mass 
terms, both Hermitian and non-Hermitian, is impossible. 

So the spontaneous emergence of non-Hermiticity in the initially Hermitian (2+1)-D GN model is 
allowed only in the chiral limit, i.e. at zero bare masses.

\section{Summary and conclusions}

In the present paper we have studied the possibility of the  dynamical appearance of both 
Hermitian and non-Hermitian mass terms in the originally Hermitian massless (2+1)-dimensional GN 
model (\ref{n1}). It is invariant with respect to several discrete transformations, two chiral 
$\Gamma^3$ and $\Gamma^5$, space reflection (or parity) $\cP$ and  time reversal $\cT$ (see in the
section \ref{IIA}). As a consequence, it is $\cP\cT$ symmetric. The problem is investigated, using a 
nonperturbative approach based on the CJT 
effective action $\Gamma(S)$ (\ref{0360}) for the composite bifermion operator 
$\bar\psi (x)\psi (y)$. In fact, $\Gamma(S)$ is a functional of a full fermion propagator $S(x,y)$ 
(see in the section \ref{IIB}). In this case, in order to find the true fermion propagator of the 
original GN model and to determine what kind of fermionic mass terms, Hermitian ${\cal M}_{H}$ or 
non-Hermitian ${\cal M}_{NH1}$ and ${\cal M}_{NH2}$ (see in Eq. (\ref{n13})), can arise dynamically 
in the model, it is sufficient to solve the stationary (gap or Dyson-Schwinger) equation (\ref{0370}) 
for the functional $\Gamma(S)$, which we have calculated up to the first order in the coupling 
constant $G$ (see Eqs. (\ref{n420}) and (\ref{n043})).  

It turns out that due to the behavior (\ref{0490}) of the bare coupling constant $G(\Lambda)$ vs 
cutoff parameter $\Lambda$, the gap equation  (\ref{n043}) is renormalized and it has three different
finite solutions. The first one (see in Sec. \ref{IIIA}) corresponds to a dynamical generation of the 
Hermitian mass term ${\cal M}_{H}$ in the model Lagrangian. In this case the phase with spontaneous 
breaking of all above mentioned discrete symmetries (including the $\cP\cT$ one) is realized in the 
system. The second finite solution of the Dyson-Schwinger equation (\ref{n043}) corresponds to 
dynamical appearing of the non-Hermitian mass term ${\cal M}_{NH1}$ in the model (see in Sec. 
\ref{IIIB}). In this case 
the phase with spontaneous emergence of non-Hermiticity is induced in the system, but it is still
$\cP\cT$ invariant and the mass spectrum of its quasiparticle excitations is real. Finally, there 
is a solution of the gap equation (\ref{n043}) 
that indicates on the possibility of spontaneous realizing in the system another non-Hermitian phase, in which 
fermionic quasiparticles are described effectively by free Lagrangian with also non-Hermitian but 
$\cP\cT$-odd mass term ${\cal M}_{NH2}$. Note that the ground state in each of these three
qualitatively different phases has the same energy density, i.e. the phases can appear spontaneously 
in the {\it massless} (2+1)-D GN model (\ref{n1}) on the same footing. It means that in the space,
filled with one of these degenerated phases, bubbles of the other two phases can be created, i.e. one 
can observe in space the mixture (or coexistence) of these three phases. Moreover, as it is noted 
in Sec. \ref{IIIB}, the effect of spontaneous non-Hermiticity can be detected only at finite $N$,
i.e. outside the large-$N$ expansion technique.

We have also shown (see in Sec. IV) that in the {\it massive} (2+1)-D GN model (\ref{n1}), i.e. when
one or another nonzero bare Hermitian mass term is added to the Lagrangian (\ref{n1}), the effect of its 
spontaneous non-Hermiticity is impossible.   

It is worth recalling that earlier the effect of spontaneous emergence of non-Hermiticity of a 
quantum system (accompanied by a real mass spectrum of its quasiparticle excitations) was discovered 
on the basis of the (3 + 1)-D NJL model \cite{Chernodub2}. Comparing our results with the results of 
this paper, we see that despite the large qualitative difference between the (2+1)-D GN  and 
(3+1)-D NJL models, the effects of their spontaneous non-Hermiticity (with real mass spectrums) have
many similar features. 
Indeed, in the NJL model, this effect is also observed only in the chiral limit, and the phase 
corresponding to it is $\cP\cT$ symmetric. In addition, the ground state of this non-Hermitian 
NJL phase has the same free energy density as some $\cP\cT$-invariant phase of the NJL model with 
Hermitian ground 
state. However, there are some differences. On the one hand, in the (2+1)-D GN model there is one 
more phase with spontaneous non-Hermiticity, which is $\cP\cT$ odd (nonsymmetric). On the other hand,
as shown in \cite{Chernodub2}, in the NJL model in the strong-coupling limit, the spontaneously 
non-Hermitian ground state generates its inhomogeneity, i.e. in this case the translational
invariance of the system is violated. Within the framework of the CJT effective action approach used 
in our paper, it is impossible to detect such a phase, since from the very beginning the presence 
of translational invariance is assumed (see in Sec. \ref{IIB}). 
  
We hope that the results of this article can be useful for describing physical phenomena in 
condensed matter systems having a planar crystal structure, or in thin films, e.g., like graphene. 
In such situations, it often happens that the elementary excitations of the system are massless. 
As a result, at low energies and in the continuum limit, physical phenomena in it can be effectively 
described by {\it massless} quantum field theory models with four-fermion interactions of the 
type (\ref{n1}) \cite{Gusynin,Ebert,Mesterhazy}. And just in these cases, the effect of spontaneous 
non-Hermiticity could be manifested.

\section{ACKNOWLEDGMENTS}

R.N.Z. is grateful for support of the Foundation for the Advancement of Theoretical Physics and 
Mathematics BASIS.

\appendix
\section{Algebra of the $\gamma$ matrices in the case of SO(2,1) group}
\label{ApC}

The two-dimensional irreducible representation of the (2+1)-dimensional
Lorentz group SO(2,1) is realized by the following $2\times 2$
$\tilde\gamma$-matrices:
\begin{eqnarray}
\tilde\gamma^0=\sigma_3=
\left (\begin{array}{cc}
1 & 0\\
0 &-1
\end{array}\right ),\,\,
\tilde\gamma^1=i\sigma_1=
\left (\begin{array}{cc}
0 & i\\
i &0
\end{array}\right ),\,\,
\tilde\gamma^2=i\sigma_2=
\left (\begin{array}{cc}
0 & 1\\
-1 &0
\end{array}\right ),
\label{C1}
\end{eqnarray}
acting on two-component Dirac spinors. They have the properties:
\begin{eqnarray}
Tr(\tilde\gamma^{\mu}\tilde\gamma^{\nu})=2g^{\mu\nu};~~
[\tilde\gamma^{\mu},\tilde\gamma^{\nu}]=-2i\varepsilon^{\mu\nu\alpha}
\tilde\gamma_{\alpha};~
~\tilde\gamma^{\mu}\tilde\gamma^{\nu}=-i\varepsilon^{\mu\nu\alpha}
\tilde\gamma_{\alpha}+g^{\mu\nu},
\label{C2}
\end{eqnarray}
where $g^{\mu\nu}=g_{\mu\nu}=diag(1,-1,-1),
~\tilde\gamma_{\alpha}=g_{\alpha\beta}\tilde\gamma^{\beta},~
\varepsilon^{012}=1$.
There is also the relation:
\begin{eqnarray}
Tr(\tilde\gamma^{\mu}\tilde\gamma^{\nu}\tilde\gamma^{\alpha})=
-2i\varepsilon^{\mu\nu\alpha}.
\label{C3}
\end{eqnarray}
Note that the definition of chiral symmetry is slightly unusual in
(2+1)-dimensions (spin is here a pseudoscalar rather than a (axial)
vector). The formal reason is simply that there exists no other $2\times 2$ matrix anticommuting with the Dirac matrices $\tilde\gamma^{\nu}$
which would allow the introduction of a $\gamma^5$-matrix in the
irreducible representation. The important concept of 'chiral'
symmetries  and their breakdown by mass terms can nevertheless be
realized also in the framework of (2+1)-dimensional quantum field
theories by considering a four-component reducible representation
for Dirac fields. In this case the Dirac spinors $\psi$ have the
following form:
\begin{eqnarray}
\psi(x)=
\left (\begin{array}{cc}
\tilde\psi_{1}(x)\\
\tilde\psi_{2}(x)
\end{array}\right ),
\label{C4}
\end{eqnarray}
with $\tilde\psi_1,\tilde\psi_2$ being two-component spinors.
In the reducible four-dimensional spinor representation one deals
with 4$\times$4 $\gamma$-matrices:
$\gamma^\mu=diag(\tilde\gamma^\mu,-\tilde\gamma^\mu)$, where
$\tilde\gamma^\mu$ are given in (\ref{C1}) (This particular reducible representation for 
$\gamma$-matrices is used, e.g., in Ref. \cite{Appelquist}). One can easily show, that
($\mu,\nu=0,1,2$):
\begin{eqnarray}
&&Tr(\gamma^\mu\gamma^\nu)=4g^{\mu\nu};~~
\gamma^\mu\gamma^\nu=\sigma^{\mu\nu}+g^{\mu\nu};~~\nonumber\\
&&\sigma^{\mu\nu}=\frac{1}{2}[\gamma^\mu,\gamma^\nu]
=diag(-i\varepsilon^{\mu\nu\alpha}\tilde\gamma_\alpha,
-i\varepsilon^{\mu\nu\alpha}\tilde\gamma_\alpha).
\label{C5}
\end{eqnarray}
In addition to the  Dirac matrices $\gamma^\mu~~(\mu=0,1,2)$ there
exist two other matrices, $\gamma^3$ and $\gamma^5$, which anticommute
with all $\gamma^\mu~~(\mu=0,1,2)$ and with themselves
\begin{eqnarray}
\gamma^3=
\left (\begin{array}{cc}
0~,& I\\
I~,& 0
\end{array}\right ),\,
\gamma^5=\gamma^0\gamma^1\gamma^2\gamma^3=
i\left (\begin{array}{cc}
0~,& -I\\
I~,& 0
\end{array}\right ),\,\,\tau=-i\gamma^3\gamma^5=
\left (\begin{array}{cc}
I~,& 0\\
0~,& -I
\end{array}\right )
\label{C6}
\end{eqnarray}
with  $I$ being the unit $2\times 2$ matrix.

\end{document}